\documentclass[12pt]{iopart}
\usepackage{iopams, graphics, epsfig}
\def\AJ{{Ap. J.} }

\def\AJS{{Ap. J. Supp.} }

\def\CQG{{\it Class. Quantum Gravity} }

\def\IJMP{{\it Int. J. Mod. Phys.} }

\def\JCP{{\it JCAP} }

\def\MNRAS{{\it Mon. Not. R. Ast. Soc.} }

\def\NP{{ Nucl. Phys.} }
\def\PL{{Phys. Lett.} }
\def\PR{{Phys. Rev.} }
\def\PRL{{Phys. Rev. Lett.} }
\def\PRTS{{\it Physics Reports} }

\begin{document}

 \title{Reconstructing K-essence}
\author{A.~A.~Sen\footnote{email: anjan.sen@vanderbilt.edu}}
\address{Department of Physics and Astronomy, Vanderbilt University,
Nashville, TN  ~~37235}

\begin{abstract}
We present a model independent method of reconstructing the Lagrangian for the k-essence field driving the present acceleration of the universe. We consider the simplest k-essence model for which the potential is constant. Later we use three parametrizations for the Hubble parameter $H(z)$, consistent with recent the SN1a data, to yield the Lagrangian $F$. Our reconstruction program does not generate any physically realistic Lagrangian for models that allow phantom crossing, whereas models without phantom crossing, yield well behaved Lagrangians.
\end{abstract} 

\pacs{98.80.-k, 98.80.Es, 98.70.Vc}

\section{Introduction}
One of the most interesting observational discoveries in the past decade has been the evidence that the expansion of the universe is speeding up rather than slowing down. A currently accelerating universe is strongly favored by type Ia supernova data \cite{super}. Recent observations of cosmic microwave background radiation (CMBR) by the Wilkinson Microwave Anisotropy Probe (WMAP) experiment \cite{wmap} and of the  large scale structure by redshift surveys e.g the Sloan Digital Sky Survey (SDSS) \cite{sdss} and the 2-Degree Field (2df) redshift survey \cite{2df} have further strengthened this result. This result together with the observational evidence that we are living in a low matter density (around $30\%$ of the critical density) flat ($\Omega_{total} =1$) universe, suggests the existence of a relatively smooth component called ``dark energy'' which dominates the energy density of the universe and also has a large negative pressure.

Although a simple cosmological constant $\Lambda$ ($\omega = -1$) can serve the purpose of this dark energy, it faces a  serious problem of fine tuning (for details, see recent review by Padamnabhan \cite{paddy}). The alternative possibility called quintessence \cite{quint}, is a varying  $\Lambda$ model, where dark energy arises from a minimally coupled scalar field rolling over its potential. For a sufficiently flat potential, the potential energy term dominates and the scalar field can mimic $\Lambda$.

Recently, an alternative possibility, that of an effective scalar field theory governed by a Lagrangian containing a non-canonical kinetic term (${\cal{L}} = - V(\phi)F(X)$, where $ X = (1/2) \partial_{\mu}\phi \partial^{\mu}\phi$) has been proposed. Such a model can lead to a late time acceleration in the present universe and is named as ``k-essence'' \cite{kes}. It is worth noting that the Generalized Chaplygin Gas model \cite{gcg} and the tachyon dark energy models \cite{tirtha} may be seen as  special cases of k-essence (e.g k-essence with a constant potential can mimic simple Chaplygin gas (CG) model \cite{frolov}).   

From the theoretical side, it is not only important to check whether dark energy is a constant or dynamical, but also if dynamical, it is equally important to know the Lagrangian for the underlying field. For a minimally coupled scalar field model, a number of studies have been done for reconstructing the potential for this scalar field in a model independent way by using observational data \cite{reconsq}. Recently Tsujikawa has studied the reconstruction of the potential function $V(\phi)$ in the k-essence Lagrangian once one assumes a certain form for the kinetic energy function $F(X)$ \cite{tsuji}. But until now, there is no attempt in the literature to reconstruct in a model independent way the function $F(X)$ in the k-essence Lagrangian. 

In this work, we propose a algorithm (similar to that proposed by Simon et al. \cite{simon} for a minimally coupled scalar field model) to reconstruct $F(X)$ in a model independent way. For present purposes, we restrict ourselves to the case where the potential $V(\phi)$ is a constant. Using this algorithm and three specific ansatzes for the Hubble parameter $H(z)$ (consistent with the supernova observations), we reconstruct  $F(X)$ in the redshift range between $z=0$ and $z=1.7$ which is the current range for the supernova data.
 
\section{Cosmological reconstruction of $F(X)$}

In a spatially flat FRW universe, the luminosity distance $D_{L}(z)$ and the coordinate distance $r(z)$ to an object at redshift $z$ are related as 

\begin{equation}
r = \int_{t}^{t_{0}} {dt\over{a(t)}} = {D_{L}(z)\over{1+z}},
\end{equation}
where we have assumed the speed of light $c =1$ and $ a_{0} = 1$, $a(t)$ is the scale factor and quantities with subscript zero denote the values at present time ($z=0$). This defines uniquely the Hubble parameter

\begin{equation}
H(z) = {\dot{a}\over{a}} = \left[{d\over{dz}}\left({D_{L}(z)\over{1+z}}\right)\right]^{-1}.
\end{equation}

This is a purely kinematic relation and does not depend on the choice of the matter part or the model of gravity. Hence knowing $D_{L}(z)$ from observations, one can unambiguously determine the Hubble parameter $H(z)$.

Next, analogous to the inflationary Hubble-flow parameter, we define the quantity \cite{simon},

\begin{equation}
\epsilon_{n+1} = {d \log |\epsilon_{n}|\over{d N}},
\end{equation}
for $n\geq 0$, where $N = \log[a(t_{i})/a(t)]$ is the number of e-foldings from some initial epoch $t_{i}$ and $\epsilon_{0} = {H_{i}\over{H}}$. One can now calculate $\epsilon_{1}$ as

\begin{equation}
\epsilon_{1} = {d\log|\epsilon_{0}|\over{d N}} = - {\dot{H}\over{H^{2}}} = q+1,
\end{equation}
where $q$ is the deceleration parameter. Similarly one can show 

\begin{equation}
\epsilon_{2} = -(1+z){d\over{dz}}[\log(q+1)],
\end{equation}
which is related to the jerk parameter $j = q + 2 q^{2} + {dq\over{dz}}$ \cite{visser}, and also $\epsilon_{3} = -(1+z){d\over{dz}}\log(\epsilon_{2})$ which is related to the snap parameter $s$ \cite{visser} (These are similiar to the ``Statefinder'' parameters \cite{state}) . We should emphasize that current supernova data can constrain up to the third derivative of the scale factor which means the deceleration parameter $q$ and the jerk parameter $j$ can be determined.

Next, we define the energy density and pressure for a k-essence fluid. For a k-field $\phi$ with a Lagrangian ${\cal{L}} = -V(\phi)F(X)$, where $X = {1\over{2}}\partial_{\mu}\phi\partial^{\mu}\phi > 0$, the energy density and pressure  are given by
\begin{eqnarray}
\rho_{\phi} &=& V[F - 2X F^{'}]\nonumber\\
p_{\phi}    &=&  -VF,
\end{eqnarray}
where $F^{'} = {dF\over{dX}}$. One can now write Einstein's equation for a flat FRW universe as
\begin{eqnarray}
3H^{2} &=& \kappa^{2}[\rho_{m} + \rho_{\phi}]\nonumber\\
{\ddot{a}\over{a}} &=& -{\kappa^{2}\over{6}}[\rho_{m}+\rho_{\phi}+3p_{\phi}],
\end{eqnarray}
where $\kappa^{2} = 8\pi G$ and  we neglect the radiation contribution as it is quite negligible around the present epoch and $\rho_{m} = \rho_{baryon}+\rho_{dm}$ is the sum of the contributions coming from baryons and dark matter.

In our subsequent calculations, we shall restrict ourselves to the simple k-essence models for which the potential $V=V_{0}$=constant. This case was first considered in \cite{kinf} and later discussed by a number of authors \cite{bob,chiba,mukh,chim,chim1,lazkoz}.  In this case one can show that 
\begin{equation}
\epsilon_{1} = {3\over{2}}{{\rho_{m}-2XF^{'}}\over{\rho_{m}+[F-2XF^{'}]}},
\end{equation}
where we have assumed $V_{0} = 1$ without any loss of generality. Also, using Einstein's equations and the expressions for $\rho_{\phi}$ and $p_{\phi}$, one can write
\begin{equation}
F = {H^{2}\over{\kappa^{2}}} [3 - 2\epsilon_{1}]
\end{equation}
Using eqns (8) and (9), one can now write
\begin{equation}
-X F^{'} = {\epsilon_{1} H^{2}\over{\kappa^{2}}} - {\rho_{m}\over{2}}.
\end{equation}

In our subsequent calculations, we shall use eqns (9) and (10) for reconstructing $F(X)$. Next using the relation $\epsilon_{1} = {(1+z)\over{H}}{dH\over{dz}}$, one can write eqn (9) as
\begin{equation}
F = {H^{2}\over{k^2}}[3 - 2(q+1)],
\end{equation}
which at the present epoch becomes
\begin{equation}
F_{0} = {H^{2}_{0}\over{\kappa^2}}[3 - 2(q_{0}+1)].
\end{equation}

Hence knowing $H(z)$ from observation, one can reconstruct $F(z)$. Also using eqn (9) one can determine $F^{'}(z)|_{0}$ (here prime denotes derivative w.r.t the argument $z$):

\begin{equation}
F^{'}(z) = {{2H^{2}}\over{k^{2}}}{\epsilon_{1}\over{1+z}}\left[3 - 2\epsilon_{1} + \epsilon_{2}\right],
\end{equation}
which gives $F(z)$ at present:

\begin{equation}
F^{'}(z)|_{0} = {2H_{0}^{2}\over{k^{2}}}[3(q_{0}+1) - 2(q_{0}+1)^{2} - ({dq\over{dz}})|_{0}].
\end{equation}
Similarly one can determine $F^{''}(z), F^{'''}(z)$ and so on at present epoch. One can always make a Taylor expansion of $F(z)$ around the present epoch $z=0$:

\begin{equation}
F(z) = F_{0} + (F^{'}|_{0}) z + {1\over{2!}} (F^{''}|_{0}) z^2 + ....
\end{equation}

\begin{figure}
\centering \resizebox{8.5cm}{!}{\includegraphics{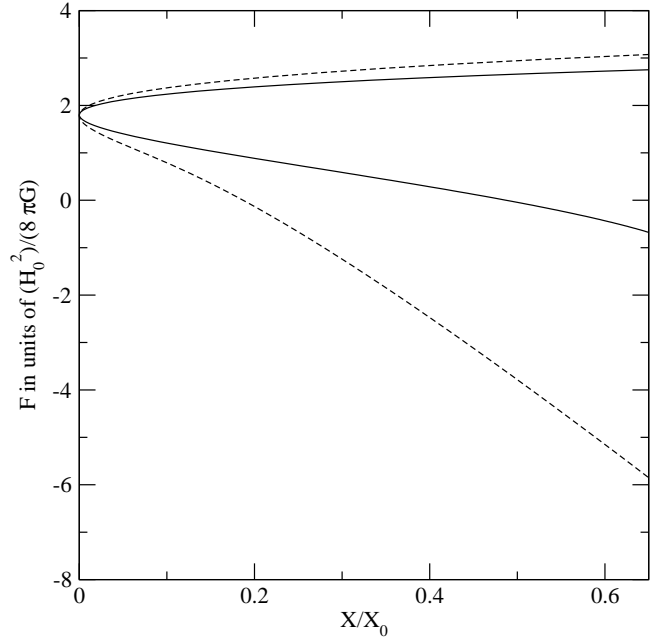}}
\caption{Reconstructed $F$ for Ansatz 1 and 2. The solid line is for ansatz 1 and the dashed line is for ansatz 2. }.
\end{figure}
Hence it is always possible to determine the approximate behavior of $F(z)$ around the present epoch by knowing $H(z)$, $q(z)$, $q'(z)$, $q^{''}(z)$ etc at present ($z=0$) from supernova observations even if one does not know the full evolution of $H(z)$. Current supernova data can only determine up to $q^{'}(z=0)$ \cite{visser}, hence in this way one can determine $F(z)$ around the present epoch up to only linear order. But as future observations start probing the universe at higher redshifts, one can determine the higher derivatives of $q(z)$, and hence the higher order behavior of $F(z)$. This provides a nice and useful method of determining the behavior of $F(z)$ around the present epoch from future distance-redshift observations.

But our goal is to reconstruct $F(X)$, which is not what eqn (11) or (15) provides. For this,  we consider the energy conservation equation for the K-Field:

\begin{equation}
(F^{'}(X) + 2 X F^{''}(X))\ddot{\phi} + 3 H F^{'}(X) \dot{\phi} = 0
\end{equation}

which gives a first integral for $F(X)$ as \cite{chim1,bob,chiba}

\begin{equation}
X F^{'2}(X) = c (1 + z)^{6},
\end{equation}
where $c$ is an arbitrary integration constant. One can now use eqn (10) to write $c$ as

\begin{equation}
\sqrt{c} = {H_{0}^{2}\over{\kappa^{2}}} {1\over{\sqrt{X_{0}}}} \left[{3\over{2}}\Omega_{m0} - \epsilon_{1}(z=0)\right],
\end{equation}
where $X_{0}$ is the present day value of $X$ and $\Omega_{m0}$ is the present day density parameter for matter. Using this expression for $c$, one can now write 
\begin{equation}
{X\over{X_{0}}} = E^{4}(z){\left[{3\over{2}}{\Omega_{m0}\over{E^{2}(z)}} - \epsilon_{1}(1+z)^{-3}\right]^{2}\over{\left[{3\over{2}}\Omega_{m0} - \epsilon_{1}(z=0)\right]^{2}}}.
\end{equation}
where $E(z) = {H\over{H_{0}}}$. Eqns (11) and (19) constitute the main equations for our reconstruction procedure. Eqn. (11) gives $F$ as a function of $z$ and Eqn. (19) gives $X$ as a function of $Z$. Hence together, they determine $F(X)$. This is important as we want to know $F(X)$ not $F(z)$. 

As mentioned above, our method of successful reconstruction of $F(X)$ depends on a sufficiently versatile fitting function for $H(z)$. For our purpose we shall use  the following flexible and model independent ansatzs:

\vspace{2mm}
\noindent
{\bf Ansatz 1}:
\begin{eqnarray}
H(z) &=& H_{0}[\Omega_{m0}(1+z)^{3}+A_{0}+A_{1}(1+z)\nonumber\\
     &+& A_{2}(1+z)^{2}]^{1/2}.
\end{eqnarray}
This has been proposed by Alam et al. \cite{alam1,alam2}. For a flat universe, the model has three independent parameters as $\Omega_{m0}+A_{0}+A_{1}+A_{2} = 1$. For $A_{1} = A_{2} =0$ the ansatz becomes an exact $\Lambda$CDM model. For $A_{0} = A_{2} = 0$ and for $A_{0} = A_{1} = 0$, it gives a dark energy model with $\omega = -2/3$ and $\omega = -1/3$ respectively. It has also been shown by Alam et al. \cite{alam1} that even for models where this ansatz is not exact, like tracker quintessence models, supergravity models or k-essence models like the Chaplygin gas, one can recover the luminosity distance to within $0.5\%$ accuracy using this ansatz in the redshift range that is relevant for supernova observations. Hence one can trust this ansatz to recover cosmological quantities to a high degree of accuracy for a large class of models. Best fit values for the parameters $A_{1}$ and $A_{2}$ for $\Omega_{m} = 0.3$ are $A_{1} = -4.16$ and $A_{2} = 1.67$, for the SN1a Gold dataset \cite{alam1} and will be used for subsequent reconstruction.

\begin{figure}
\centering \resizebox{8.5cm}{!}{\includegraphics{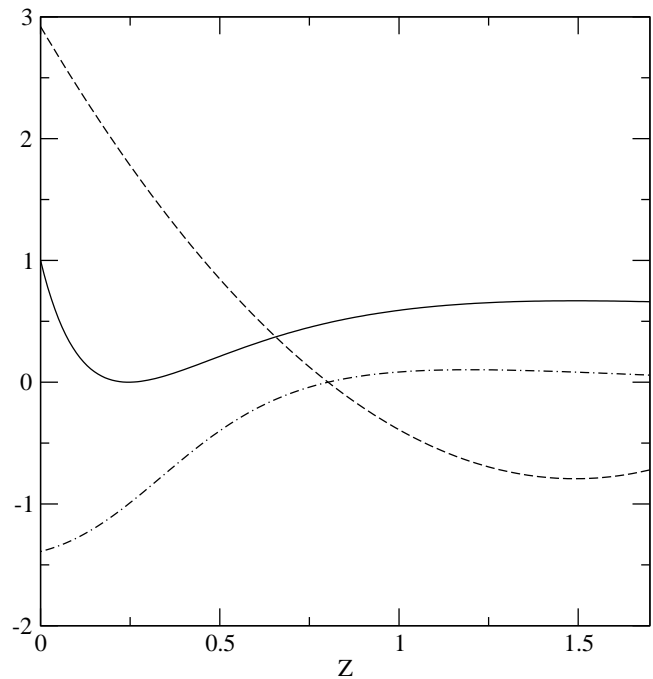}}
\caption{Variation of ${X/X_{0}}$ (solid line), $F$ (dashed line) and equation of state for k-essence field (dot-dashed line) for Ansatz 1}.
\end{figure}
 
\vspace{50mm}
\noindent
{\bf Ansatz 2}:
\begin{equation}
w(z) = w_{0} + {w_{1}z\over{1+z}},
\end{equation}
where $w_{0}$ and $w_{1}$ are constants. This ansatz was first discussed by Chevallier and Polarski \cite{pol} and later studied more elaborately  by Linder \cite{lind}. For low redshifts, the ansatz reduces to the conventional first order expansion to the equation of state $w = w_{0} + w_{1}z$. It is also well behaved and bounded for high redshifts and can mimic several scalar field equations of state.  For example, for the supergravity model, the ansatz can mimic the distance-redshift behavior to $0.2\%$ up to redshift $z \sim 1100$. The best fit values for this model while fitting with the SN1a gold dataset are $w_{0} = -1.58$ and $w_{1} = 3.29$ for $\Omega_{m} = 0.3$ \cite{alam1}. We shall use these values for the parameters for the reconstruction.

\vspace{2mm}
\noindent
{\bf Ansatz 3}:
\begin{eqnarray}
H(z) = H_{0}[\Omega_{m0}(1+z)^3 \nonumber\\
+ (1-\Omega_{m0})(A_{s}+(1-A_{s})(1+z)^{3(1+\alpha)})^{1/(1+\alpha)}]^{1/2}
\end{eqnarray} 
where $A_{s}$ and $\alpha$ are constant. This ansatz for the Hubble parameter $H(z)$ can be obtained by modeling the dark energy as Generalized Chaplygin gas with an equation of state $p_{ch} = -{A\over{\rho_{ch}^{\alpha}}}$ \cite{gcg}. By choosing different ranges for the parameters  $A_{s}$ and $\alpha$, this ansatz can behave as standard dark energy model with asymptotic de-Sitter phase, early phantom model, late phantom phantom model as well as transient model where the present acceleration of the universe is only temporary as it again enters the dust-dominated decelerating phase in future. Recently it has been shown that current SN1a gold data can allow all four behaviors of this model with suitable choice of the matter density parameter $\Omega_{m0}$ \cite{ggcg}. For our analysis, we have considered four combinations of ($A_{s}, \alpha$): (0.95, 0.2), (0.85, -1.2), (1.1 -0.8) and (1.5, -1.7), which result in the standard, transient, early phantom and late phantom phases respectively for $\Omega_{m0}$ = 0.2, 0.25, 0.3 and 0.35. All these combinations for ($A_{s}, \alpha$) are within the $1\sigma$ confidence contour for SN1a Gold dataset \cite{ggcg}.
\begin{figure}
\centering \resizebox{8.5cm}{!}{\includegraphics{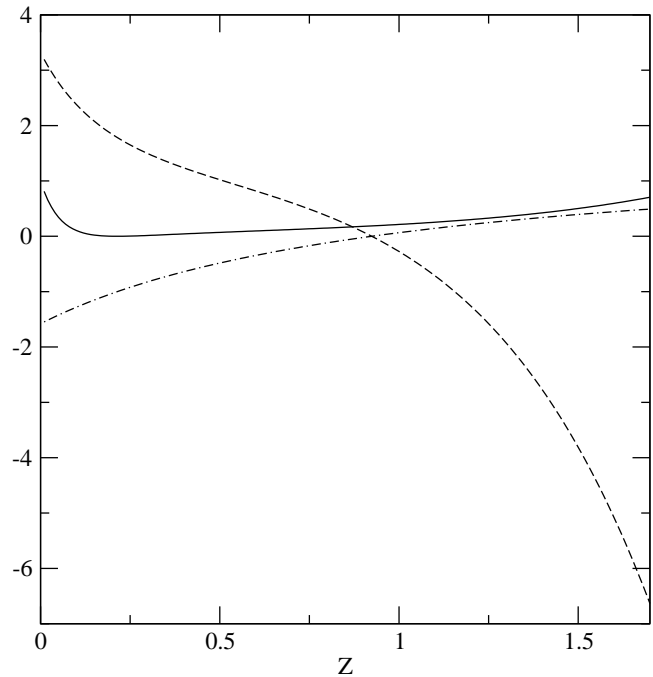}}
\caption{Variation of ${X/X_{0}}$ (solid line), $F$ (dashed line)
 and equation of state for k-essence field (dot-dashed line) for Ansatz 2}.
\end{figure}

\vspace{2mm}
\noindent
{\bf Results}:

\vspace{2mm}
\noindent In Fig 1, we have plotted the reconstructed $F$ as a function of ${X/X_{0}}$ for ansatz 1 and 2. For both of these ansatzes, the reconstructed F(X) is a double-valued function of X. This behaviour of $F(X)$ is not what one should expect for any physically realistic Lagrangian. To see the reason behind this pathological behaviour for $F(X)$, one has to carefully study the equations. First of all, one can see both of these ansatzes allow the crossing of the phantom divide ($\omega= -1$) \cite{phantom}. Also $X(z)$ has a minimum in both cases with $X_{min} = 0$ and the minimum occurs precisely when the equation of state crosses the phantom divide $\omega = -1$ (see Fig 2 and 3). On the other hand, $F$ is a monotonically decreasing function with $z$ within the relevant redshift range. This forces $F$ to be a double-valued function of $X$ when $F$ doubles back on itself at $X=0$. Hence one might suspect that this unphysical behaviour of $F(X)$ is due to the crossing of phantom divide.      
\begin{figure}
\centering \resizebox{8.5cm}{!}{\includegraphics{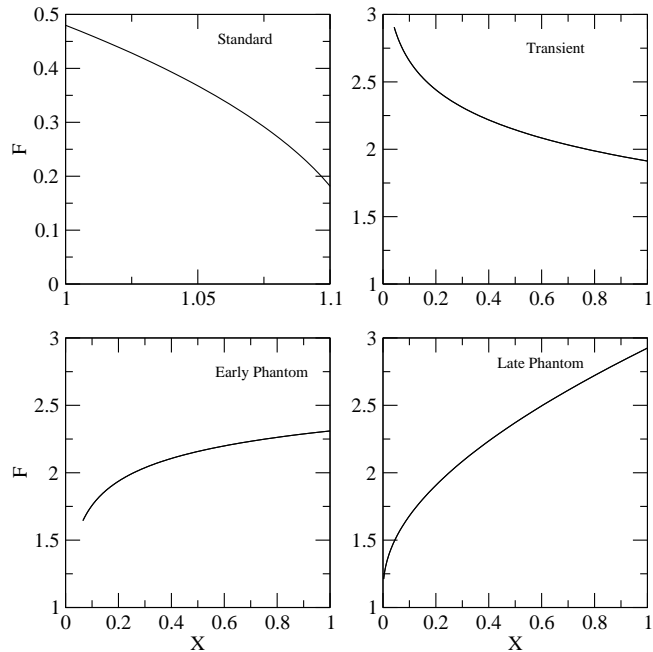}}
\caption{Reconstructed F(X) for  anstaz 3 for four different cases (See text)}.
\end{figure}
To see that this is actually the reason, let us consider the equation of state of the k-essence fluid:
\begin{equation}
\omega = {F(X)\over{2X F^{'}(X) -F(X)}}.
\end{equation} 

\noindent
In order to have $\omega = -1$, one should have $X=0$, or $F^{'}(X)=0$, or both. But looking at the equation of motion (eqn 17), one concludes that it is not possible (except at future infinity $z = -1$), unless $X=0$ and $F^{'}(X) \rightarrow \infty$ as ${1\over{\sqrt{X}}}$, so that the l.h.s of eqn. 17 remains finite. Also, to cross the boundary $\omega = -1$, $F^{'}(X)$ should change its sign at $X=0$. These two conditions (e.g $F^{'}(X)$ diverges and changes sign at $X=0$) together single out the sole possibility of $F(X)$ being a double-valued function  and crossing back on itself at $X=0$, the point of phantom crossing. This rules out any physically realistic Lagrangian $F(X)$ for a k-essence field which allows phantom crossing. This is consistent with earlier results by Vikman \cite{vik} and Caldwell and Doran \cite{doran}  who have made a more elaborate study for phantom crossing with a k-essence field and arrived at similar  conclusions. (see also the more recent investigation \cite{neto}.)

But one can check that restricting the equation of state from crossing the phantom divide can result in a  well-behaved reconstructed $F(X)$. This happens for our ansatz 3 which although perfectly consistent with the recent SN1a data, never crosses the phantom divide. In this case the reconstructed $F(X)$ is shown in Fig 4 for four different models and it can be seen that $F(X)$ is a perfectly well-behaved single valued function for all four cases. It is also interesting to note that for standard and transient models (non-phantom cases), $F$ is a monotonically decresing function of $X$ whereas for early and late phantom models, it is an increasing function of $X$ (see Fig. 4). This is because for $X>0$, the sign of ${F^{'}(X)\over{F(X)}}$ is related to the value of $\omega$ (see Eqn.(23)) and for phantom models, $F^{'}(X) >0$, whereas for non-phantom case, $F^{'}(X) <0$ as $F(X)$ is always positive for all four cases in the relevant redshift range (see Fig 5).

The behaviors of $X(z)$ and $F(z)$ for this ansatz are shown in Fig 5.

\begin{figure}
\centering \resizebox{8.5cm}{!}{\includegraphics{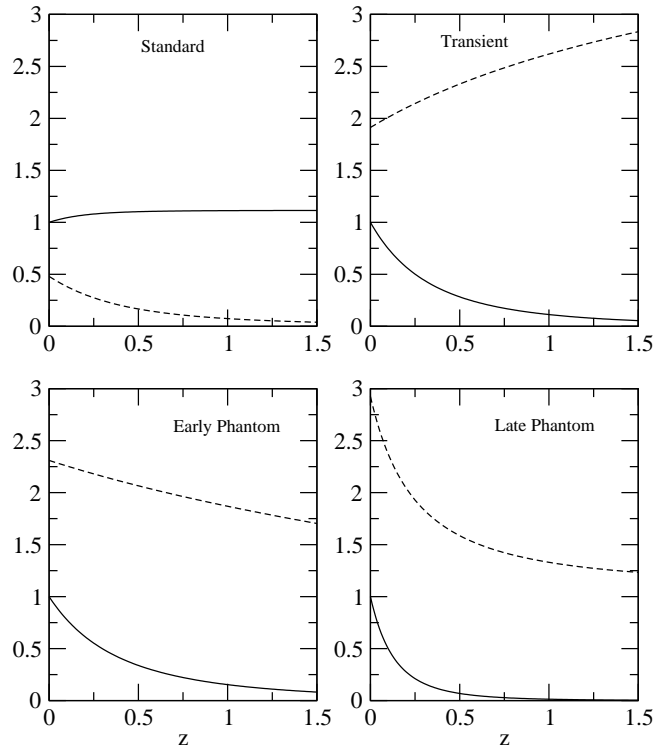}}
\caption{Variation of X(z) (solid line) and F(z) (dashed line) for four different cases (See Text) for ansatz 3}.
\end{figure}
 
\section{Conclusion}
In this work, we have proposed a method for reconstructing the k-essence Lagrangian for constant potential. In doing so, we have defined the parameters $\epsilon_{n}$, which are similar to the horizon-flow parameters for inflation. Using these parameters, we have related $F(z)$ and $F'(z)$ to the directly observable quantities like $\Omega_{m}$, $q(z)$, $H(z)$. Equations (11) and (19) are the core of our reconstruction program. By knowing $H(z)$ from distance-redshift measurements like SNIa observations, one can reconstruct $F$ as a function of redshift $z$ and $X$, the kinetic energy. Later, we have used three well known ansatz for $H(z)$, in order to reconstruct $F$ for $z=0$ to $z=1.7$ which is the current redshift range for SNIa data. We should emphasize that our reconstruction program only involves that portion of $F(X)$ over which the field evolves to give the required $H(z)$ and is insensitive to certain ranges in $X$.

Our study shows that in order to have a physically realistic k-essence Lagrangian $F(X)$, the crossing of the phantom boundary $\omega = -1$ is forbidden. For ansatzes 1 and 2, which allow the phantom crossing, the reconstructed $F$'s are physically unrealistic double-valued functions of $X$. This result is consistent with earlier investigations with similar conclusion \cite{vik, doran, neto}. For ansatz 3, which is also consistent with SN1a data but does not allow phantom crossing, the reconstructed $F(X)$ is a well-behaved, single valued function. The result also shows that one has to be careful while choosing different ansatzes for $H(z)$ or $\omega(z)$ for dark energy to fit different observational data. While some of them are perfectly consistent with observation, it may not be possible to relate them to any physically realistic theory. 

\section{Acknowledgment}
I would like to  thank Robert J. Scherrer for carefully reading the manuscript and providing important suggestions and comments. I also thank M. Doran for helpful discussions. The author is also grateful to the anonymous referee pointing one crucial error in Fig 2 and Fig 3.

\end{document}